\documentclass[%
 aip,
 amsmath,amssymb,
 reprint,%
]{revtex4-1}

\usepackage{graphicx}
\usepackage{dcolumn}
\usepackage{bm}

\usepackage[utf8]{inputenc}
\usepackage[T1]{fontenc}
\usepackage{mathptmx}
\usepackage{etoolbox}

\usepackage{xcolor}

\makeatletter
\def\@email#1#2{%
 \endgroup
 \patchcmd{\titleblock@produce}
  {\frontmatter@RRAPformat}
  {\frontmatter@RRAPformat{\produce@RRAP{*#1\href{mailto:#2}{#2}}}\frontmatter@RRAPformat}
  {}{}
}%
\makeatother
\begin{document}

\preprint{AIP/123-QED}

\title[]{
{\color{black} Quantum Mpemba effect from initial system-reservoir entanglement}}
\author{Stefano Longhi}
 \email{stefano.longhi@polimi.it}
 \altaffiliation[Also at ]{IFISC (UIB-CSIC), Instituto de Fisica Interdisciplinar y Sistemas Complejos - Palma de Mallorca, Spain}
\affiliation{ Dipartimento di Fisica, Politecnico di Milano, Piazza L. da Vinci 32, I-20133 Milano, Italy
}%


\date{\today}

\begin{abstract}
The Mpemba effect--where hot systems cool faster than colder ones-- has intrigued both classical and quantum thermodynamics. As compared to classical systems, quantum systems add complexity due to quantum correlations. Recent works have explored anomalous relaxation and Mpemba-like effects in several quantum systems, considering isolated systems at zero temperature or open systems in contact with reservoirs under Markovian or non-Markovian dynamics. However, these models typically assume an initial unentangled system-bath state, overlooking the role of initial system-environment correlations.  Here  we propose a type of quantum Mpemba effect, distinct from the strong Mpemba effect, originating from initial system-bath entanglement solely. It is shown that the degree of initial entanglement significantly influences the early relaxation dynamics, with certain conditions causing backflow and retarded thermalization. As an example, we investigate the spontaneous emission of a two-level atom in a photonic waveguide at zero temperature, where an initial atom-photon entangled state results in delayed relaxation and pronounced Mpemba effect.
These findings highlight the crucial role of quantum correlations in thermalization processes and open new avenues for identifying and engineering quantum Mpemba phenomena. Controlling relaxation dynamics through system-environment entanglement may have potential applications in quantum thermal machines, state initialization protocols, and quantum information processing, where precise control over thermalization is essential.
\end{abstract}
\maketitle

The Mpemba effect (ME), where hot systems cool faster than colder ones under certain conditions, has long intrigued both scientists and laypeople alike \cite{S1,S2,S3,S4,review}. First observed in the 1960s \cite{S1}, this counterintuitive phenomenon has traditionally been studied within the realm of classical thermodynamics (see e.g. \cite{S4,S5,S6,S7,S8,S9,S10,S11,S12} and references therein). While the effect's exact mechanisms remain debated \cite{S4,S6}, its persistence across various systems and conditions underscores the complexity of thermal relaxation processes.
The prominent role of memory effects in the ME has been highlighted on several occasions-- particularly in complex spin-glass models, 
where the ME is essentially a nonequilibrium memory effect, encoded in the
glassy coherence length \cite{Referee}. In recent years, attention has turned to quantum systems, where the interplay between quantum correlations, system-bath interactions, and thermalization dynamics adds layers of complexity to the ME \cite{Q1,Q2,Q3,Q4,Q5,Q6,Q7,Q8,Q9,Q10,Q11,Q12,Q13,Q14,Q15,Q16,Q17,Q18,Q19,Q20,Q21,Q22,Q23,Q24,Q25,Q26,Q26b,Q27,Q28,review2} . Several recent studies highlighted the anomalous thermalization behaviors exhibited by certain quantum systems considering  isolated systems at zero temperature \cite{Q7,Q10,Q11,Q22} or open systems in contact with thermal reservoirs under various types of dynamics \cite{Q1,Q2,Q3,Q4,Q8,Q13,Q14,Q15,Q18}, such as Markovian \cite{Q3,Q4,Q8,Q13,Q15} and non-Markovian \cite{Q23}. 
For thorough and up-to-date reviews of ME and its multifaceted aspects in both classical and quantum systems, refer to \cite{review,review2}.  In many of these models, it has been assumed that the system and the bath are in a disentangled state at the outset, i.e. in a product state with the bath at equilibrium, which has led to the neglect of initial system-bath correlations in the analysis of relaxation dynamics. However, it is precisely these correlations that are expected to play a pivotal role in the way a system exchanges energy with its environment \cite{C1,C2,C3,C4,C5,C6,C7,C8,C9,C10,C11,C12,C13,C14}.
When the system and the environment are initially correlated,
the situation is more complicated, and it is still an open
problem to understand the relationship between the structure
of the initial system-bath states and the nature of the resulting
dynamics \cite{C10,C11}.\\
In this work, we introduce a novel type of quantum Mpemba effect that emerges from initial system-bath entanglement-- a mechanism fundamentally distinct from previously studied scenarios, which typically involve accelerated thermalization (commonly referred to as the strong Mpemba effect) or rely on non-Markovian dynamics. We show that entanglement present at the onset of thermal relaxation can profoundly influence the system's subsequent evolution. In particular, we find that the degree of initial entanglement can significantly alter early-time relaxation behavior, leading to cases where a system initially further from equilibrium relaxes faster than one closer to it.
To illustrate this phenomenon, we examine the spontaneous emission of a two-level atom coupled to a photonic waveguide at zero temperature. When the atom and photon field are initially unentangled, with the field in the vacuum state, the atom exhibits the typical nearly-exponential decay expected in the weak-coupling regime and for a flat continuum of electromagnetic modes. In contrast, when the atom and photon field begin in an entangled state corresponding to a time-reversed decay, the emission is noticeably delayed--providing a clear and striking example of a quantum Mpemba effect driven by initial entanglement.


The idea of ME induced by initial system-reservoir correlations is illustrated in Fig.1. Let us consider a bipartite isolated quantum system consisting of a finite system S coupled to a macroscopically large reservoir (such as a thermal bath) B, described by the Hamiltonian $H=H_S+H_B+V$, where $V$ is the interaction Hamiltonian. We denote by $\rho(t)$ and $\rho_S(t)={\rm Tr}_B (\rho(t))$ the density operator of the full bipartite system and reduced density operator of S, respectively, where Tr$_B$ denotes the partial trace over the reservoir. For an initial  state $\rho(0)$ of the composite system, the reduced density operator of S at subsequent times $t$ can then be written as
\begin{equation}
\rho_S(t)={\rm Tr}_B(U_t \rho(0)U^{\dag}_t), 
\end{equation}
where $U_t=\exp(-iHt)$ is the unitary time evolution operator of the composite system. If one assumes that the system and the bath are initially uncorrelated with a fixed bath state
$\rho_B$, given by the Gibbs state at temperature $T$, i. e., $\rho(0)=\rho_S(0) \otimes \rho_B$ and $\rho_B=\exp(-H_B/k_BT) / {\rm Tr} (\exp(-H_B/k_BT))$, one can describe the time evolution of the reduced system $\rho_S(t)$ through a family of completely positive dynamical maps that are contractive for the trace distance \cite{Petruccione}. In the weak-coupling limit  at leading order one has \cite{Petruccione} $\rho(t) \simeq \rho_s(t) \otimes \rho_B$, i.e. system-bath correlations and entanglement arising from the dynamics are perturbative corrections. Let us assume 
that for rather arbitrary initial system state  $\rho_{S}(0)$ the reduced density operator $\rho_S(t)$ reaches a unique stationary (equilibrium) state $\rho_E$ as $t \rightarrow \infty$. This equilibrium state is expected to be the thermal (Gibbs) state $\rho_E=\exp(-H_S/k_BT) / {\rm Tr} (\exp(-H_S/k_BT))$ at the temperature $T$ of the bath (see e.g. Sec.3.3.2 of Ref.\cite{Petruccione}). Further, since the time evolution of the reduced system $\rho_S(t)$ is described by a completely positive dynamical map that is contractive for the trace distance, for any time $ t \geq 0$ one has \cite{C2}
\begin{equation}
D(\rho_S(t))  \leq D(\rho_S(0))  
\end{equation}
and $\lim_{t \rightarrow \infty} D(\rho_S(t))=0$,
where $D(\rho_S(t))=(1/2){\rm Tr}(|\rho_S(t)-\rho_E|)$ is the trace distance between the state $\rho_S(t)$ and the equilibrium state $\rho_E$. 
The ME corresponds to the situation where, for two initial states $\rho_S^{(1)}(0)$ and $\rho_S^{(2)}(0)$, such that $D(\rho_S^{(1)}(0)) > D(\rho_S^{(2)}(0))$, asymptotically for $t \rightarrow \infty$ one has $D(\rho_S^{(2)}(t)) > D(\rho_S^{(1)}(t))$, i.e. the initial state 1 farther from equilibrium reaches faster the equilibrium state than the other state 2 initially closer to equilibrium. The typical situation is where the evolution of $\rho_S(t)$ is governed by a  master equation of Lindblad form with a Liouvillian superoperator $\mathcal{L}$, i.e. $(d \rho_S/dt)= \mathcal{L} \rho_S$, such that the relaxation dynamics is fully embodied in the left/right eigenoperators and eigenvalues of the Liouvillian. In this case the ME (also dubbed the "strong'' ME) is generally observed when state $\rho_S^{(1)}(0)$ is orthogonal to the slowest decaying left eigenoperator of $\mathcal{L}$, thus displaying an accelerated relaxation as compared to state $\rho_S^{(2)}(0)$ (see e.g. \cite{review,Q3,Q13,Q15}). An illustration of the strong ME is depicted in Fig.1(a).\\ 
\begin{figure}
\includegraphics[width=8 cm]{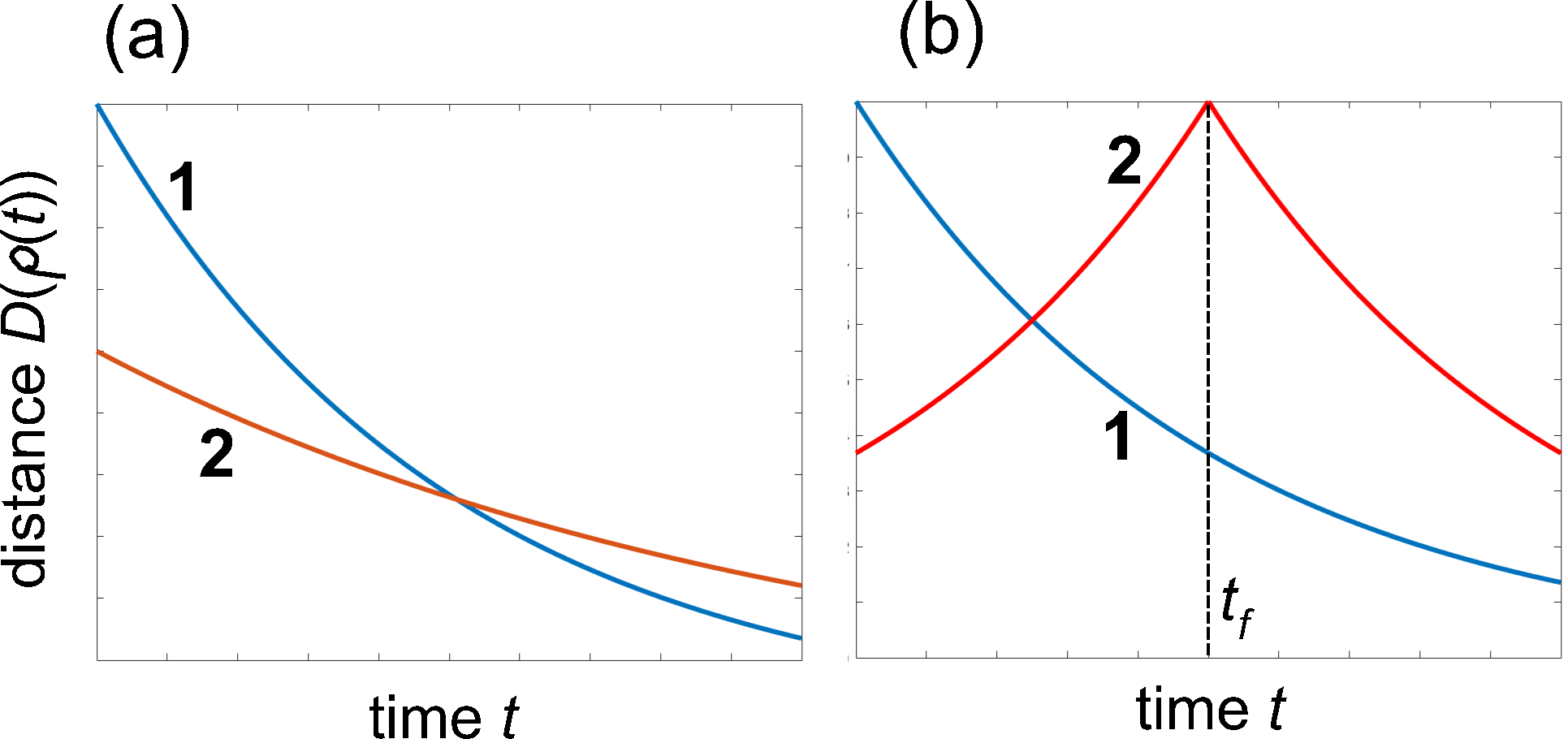}
\caption{Schematic of the two kinds of quantum Mpemba effect arising (a) from different relaxation rates of states 1 and 2 (strong Mpemba effect), and (b) from an initial entangled system-bath state.  In (a) the initial state of the composite system is the canonical product state $\rho(0)=\rho_S(0) \otimes \rho_B$, where $\rho_B$ is the Gibbs state of the bath at temperature $T$. The two different initial states 1 and 2 of the system, with density operators $\rho_S^{(1)}(0)$ and $\rho_S^{(2)}(0)$, display different decay rates. In (b) state 1 is the same as in (a), corresponding to an initial uncorrelated product state of the composite system, while 2 corresponds to an initial entangled system-bath state, as discussed in the main text.}
\end{figure}
However, we may break the main assumption that at initial time the system-bath state $\rho(0)$ is strictly an uncorrelated product state with the bath at thermal equilibrium. 
For example, we may consider a more general initial condition $\rho(0)$  corresponding to a correlated or an entangled system-bath state such that the bath state $\rho_B(0)={\rm Tr}_S(\rho(0))$ is perturbed from the the Gibbs state. In this case, $\rho_S(t)$ still relaxes to the same equilibrium state $\rho_E$ as in the the previous case. However, Eq.(2) can be violated and, owing to transient backflow of information from the bath to the system,  the distance from equilibrium can even increase transiently \cite{C2,C3}. This behavior clearly facilitates the appearance of ME since the relaxation to equilibrium can be strongly modified and delayed by initial system-bath correlation or entanglement.  To illustrate this point, let us compare the relaxation behaviors in the weak coupling regime under two different initial conditions. In the first case the initial state of the composite system is given by the canonical product state $\rho^{(1)}(0)=\rho_S^{(1)}(0) \otimes \rho_B^{(1)}(0)$, 
where the bath state $\rho_B^{(1)}(0)$ is the Gibbs equilibrium state at temperature $T$, $\rho_B^{(1)}(0)=\exp(-H_B / k_BT) / {\rm Tr} (\exp(-H_B / k_BT))$. 
Owing to the system-bath coupling, at subsequent times $\rho^{(1)}(t)$ is not strictly a product state and the system becomes entangled with the bath. The reduced density operator $\rho_S^{(1)}(t)={\rm Tr}_B (\rho^{(1)}(t))$ relaxes toward $\rho_E$,  given by the Gibbs state $\rho_E=\exp(-H_S / k_BT) / {\rm Tr} (\exp(-H_S / k_BT))$, with a distance $D(\rho_S^{(1)}(t))$ that satisfies inequality (2) and vanishes as $ t \rightarrow \infty$. In particular, after some long enough time $t=t_f$, of the order or larger than the typical relaxation time,  the system has almost reached the final equilibrium state $\rho_E$, and thus $D(\rho_S^{(1)}(t_f)) \simeq 0$. Let us now consider another case: at initial time $t=0$ the system-bath state is described by the entangled state $ \rho^{(2)}(0)=U_{-t_f} \rho^{(1)}(0) U_{-t_f}^{\dag}$. Basically, $\rho^{(2)}(0)$ is the time-reversed evolved state, for a time $t_f$, of the previously considered canonical product state initial condition.  Considering the time reversal symmetry exhibited by microscopic dynamics \cite{revers}, we expect $\rho_S^{(2)}(0)=\rho_S^{(1)}(-t_f)$ to be close  to the equilibrium Gibbs state $\rho_E$, i.e. 
 one has $D(\rho_S^{(2)}(0)) \simeq 0$. This happens when the full Hamiltonian $H$ displays time-reversal symmetry and $\rho_S^{(1)}(0)$ is an Hermitian symmetric matrix in the energy eigenbasis of $H_S$,  as shown in the Appendix A. In the subsequent time evolution, there is a backflow from the bath to the system, and at $t=t_f$ one has exactly $\rho^{(2)}(t_f)=U_{t_f} \rho^{(2)}(0) U^{\dag}_{t_f}= 
  \rho^{(1)}(0)$. This means that for $t>t_f$ one has $\rho^{(2)}(t)= \rho^{(1)}(t-t_f)$ and thus $D(\rho_S^{(2)}(t))=D(\rho_S^{(1)}(t-t_f))$, indicating that the relaxation process toward the equilibrium $\rho_E$ is delayed, as compared to the previous case, by the time $t_f$. Since $t_f$ can be taken arbitrarily long, a pronounced ME is expected to be observed; see Fig.1(b) for a sketch. Note that, since the asymptotic temporal decay of the distance is the same for the two states $\rho_S^{(1)}(t)$ and $\rho_S^{(2)}(t)$, this kind of ME is not a strong ME, i.e. it does not result from a larger relaxation rate of one of the two states, rather from the initial different transient dynamics. \par   
To illustrate  the emergence of quantum ME from initial system-bath entanglement and its distinctive feature from the strong ME, let us consider the spontaneous emission of a quantum emitter in contact with a one-dimensional bosonic bath \cite{Petruccione}, which can describe for example the spontaneous emission process of a two-level atom in a waveguide or in a lossy cavity \cite{SP1,SP2,SP2b,SP2c,SP3,SP4,SP5} or of an artificial atom in circuit quantum electrodynamics \cite{CQED1,CQED2,CQED3}. Let us indicate by $|e\rangle$ and $|g \rangle$ the excited and ground states of the atom, with transition frequency $\omega_0$, and let assume that the atom is electric-dipole coupled with a continuum of electromagnetic modes of the waveguide, each with associated wave number $k$,
dispersion relation $\omega=\omega(k)$ with $\omega(-k)=\omega(k)$ and annihilation (creation) operator $a_k$ ($a^{\dag}_k$), satisfying the usual bosonic commutation relations $[a_k,a_{k'}]=[a^{\dag}_k,a^{\dag}_{k'}]=0$ and $[a_k,a^{\dag}_{k'}]=\delta(k-k')$. In the rotating-wave approximation, the Hamiltonian of the atom-photon field reads 
\begin{equation}
H=\omega_0 |e \rangle \langle e|+ \int dk \omega(k) a^{\dag}_ka_k+ \int dk \left\{  g(k) a_k |e \rangle \langle g|+{\rm H.c.} \right\}
\end{equation}
where $g(k)$ is the coupling constant, with $g(-k)=g(k)$ for non-chiral waveguides. The Hamiltonian displays time reversal symmetry when the coupling $g(k)$ is real. The explicit expressions of the dispersion relation $\omega(k)$ and spectral coupling factor $g(k)$ depend on the specific waveguide model.
For example, if we consider an atom coupled to an array of optical resonators \cite{CROW1,CROW2,CROW3,CROW4}, as schematically shown in Fig.2(a), the full Hamiltonian of the system reads \cite{}
\begin{eqnarray}
H & = & \omega_0 |e \rangle \langle e|+ \sum_l  \left\{\omega_c c^{\dag}_l c_l-J(c^{\dag}_{l+1}c_l+{\rm H.c.}) \right\} \nonumber \\
&+ &g_0 \left( c_0^{\dag} |g \rangle \langle e| +{\rm H.c.} \right)
\end{eqnarray}
where $\omega_c \simeq \omega_0$ is the resonance frequency of the electromagnetic mode  in each single resonator, $J$ is the photon hopping rate between adjacent resonators, $c^{\dag}_l$ ($c_l$) is the creation (destruction) operator of the photon field in the $l$-th resonator of the array (Wannier basis), and $g_0$ is the electric-dipole coupling strength of the two-level atom placed in the $l=0$ cavity. The Hamiltonian (4) can be cast in the canonical form Eq.(3), with $\omega(k)=\omega_c-2J \cos(k)$ and $g(k)=g_0/ \sqrt{2 \pi}$, after introduction of the creation/destruction operators of Bloch modes of the photonic lattice, i.e. after letting $a(k)=(1/ \sqrt{2 \pi}) \sum_l c_l \exp(ikl)$, where $-\pi \leq k < \pi$ is the Bloch wave number.\\ 
Since $H$ commutes with $N=|e \rangle \langle e| + \int dk a^{\dag}_k a_k$, the total number of excitations $N$ is conserved. Assuming that the photon field is at zero temperature, to study the process of spontaneous emission one can restrict the analysis considering the subspace with excitations $N=1$ solely (see e.g. \cite{SP2b,SP3,SP4}). Assuming for the sake of definiteness that the atom-photon system is initially in a pure state, in the single excitation subspace the density operator is given by $\rho(t)= | \psi(t) \rangle \langle \psi(t)|$ with the wave function
\begin{equation}
|\psi(t) \rangle= \exp(-i \omega_0 t) \left\{ c_a(t) |e \rangle \otimes |0 \rangle + \int dk \varphi(k,t) |g \rangle \otimes a^{\dag}_k |0 \rangle \right\}
\end{equation}
where $|0 \rangle$ is the vacuum state of the photon field, $c_a(t)$ is the amplitude probability to find the atom in the excited state at time $t$, and $\varphi(k,t)$ is the amplitude probability that a photon has been emitted at time $t$ in the $k$-mode, with $|c_a(t)|^2+\int dk |\varphi(k,t)|^2=1$. The amplitude probabilities satisfy the coupled equations
\begin{eqnarray}
i \frac{dc_a}{dt} & = & \int dk g(k) \varphi(k,t) \\
i \frac{\partial \varphi}{\partial t} & = & [\omega(k)-\omega_0] \varphi(k,t)+ g^*(k) c_a(t) 
\end{eqnarray}
with given initial conditions.
The reduced density operator for the atom, $\rho_S(t)={\rm Tr}_B (\rho(t))$, is diagonal in the $\{ |g\rangle, |e \rangle \}$ basis and reads $\rho_S(t)=|c_a(t)|^2 | e \rangle \langle e|+(1-|c_a(t)|^2) |g \rangle \langle g|$,
i.e. the survival probability $|c_a(t)|^2$  is the only dynamical variable of the system. For a featureless continuum and in the absence of atom-photon bound states, $c_a(t) \rightarrow 0$ as $t \rightarrow \infty$ for arbitrary initial conditions, i.e. the equilibrium state is $\rho_E=|g \rangle \langle g|$, which is indeed the Gibbs distribution at zero temperature. The distance $D(\rho_S(t))$ from equilibrium is simply given by $D(\rho_S(t))=|c_a(t)|^2$. It is clear, however, that the spontaneous emission process is strongly affected by the initial system-bath state, i.e. by the values $c_a(0)$ and $\varphi(k,0)$, even when we operate in the weak-coupling regime.\\
  Let us indicate by $A(t)=c_a(t)$ and $\Phi(k,t)=\varphi(k,t)$ the solution to Eqs.(6) and (7) corresponding to the canonical initial condition $c_a(0)=1$ and $\varphi(k,0)=0$, i.e. when the atom is initially in the excited state $|e \rangle$ and the photon field is in the vacuum state (see e.g. \cite{SP2b}). This initial canonical condition corresponds to the initial product state $\rho^{(1)}(0)=\rho_S^{(1)}(0) \otimes \rho_B^{(1)}(0)$, where $\rho_S^{(1)}(0)=|e \rangle \langle e|$ and $\rho_B^{(1)}=|0 \rangle \langle 0|$ is the Gibbs state of the photon field at zero temperature. The amplitude $A(t)$ satisfies an integro-differential equation that accounts for memory effects. Under the usual weak-coupling and Markov approximations, i.e. when the memory time $\tau_m \rightarrow 0$, $A(t)$ decays exponentially in time, i.e. $|A(t)|^2=\exp(- \Gamma t)$. The decay rate $\Gamma$ is given by the usual Fermi golden rule and reads $\Gamma=4 \pi |g(k_0)|^2/v_g$, where $k_0$ is the wave number such that $\omega( \pm k_0)=\omega_0$ and $v_g=| (d \omega/dk)_{k_0}|$ is the group velocity of the photon field at frequency $\omega_0$ ( 
  technical details are given in Appendix B). Therefore, if we start with the canonical initial condition $\rho^{(1)}(0)=|e \rangle \langle e| \otimes |0 \rangle \langle 0|$, the relaxation $D(\rho_S^{(1)}(t))$ toward equilibrium is given by the usual exponential decay law $D(\rho_S^{(1)}(t))=\exp(-\Gamma t)$ of spontaneous emission. 
  As an example, curve 1 in Fig.2(b) shows the exact decay behavior of $D({\rho^{(1)}_S(t)})=|c_a(t)|^2$ for the coupled-cavity waveguide lattice of Fig.2(a) for parameter values $\omega_c=\omega_0$ and $g_0/J=0.2$. The decay curve  has been numerically computed solving the dynamical equations for the Hamiltonian (4) in Wannier basis assuming the canonical initial condition. The curve is well fitted by the predicted exponential profile with a decay rate $\Gamma=g_0^2/J$ given by the Fermi golden rule, indicated by the dashed curve in Fig.2(b).\\
Let us now consider a different initial condition, where the atom and the photon field are entangled, namely let us assume $c_a(0)=A^*(t_f)$ and $\varphi(k,0)=\Phi^*(k,t_f)$, i.e. $\rho^{(2)}(0)=| \psi(0) \rangle \langle \psi(0)|$ with
\begin{equation}
|\psi(0) \rangle=A^*(t_f) |e \rangle \otimes |0 \rangle+ |g \rangle \otimes \int dk \Phi^*(k,t_f) a^{\dag}_k |0 \rangle
\end{equation} 
where $t_f$ is a fixed time, of the order or larger than the spontaneous decay time $1/ \Gamma$. Note that in this case $\rho_S^{(2)}(0)=|A(t_f)|^2=\exp(- \Gamma t_f /2)$, which is significantly smaller than 1.  Under time reversal symmetry, i.e. $g(k)$ real, it readily follows that, for $0 \leq  t \leq t_f$ the solutions to Eqs.(6) and (7) with the initial condition $c_a(0)=A^*(t_f)$ and  $\varphi(k,0)=\Phi^*(k,t_f)$ is given by $c_a(t)=A^*(t_f-t)$ and $\varphi(k,t)=\Phi^*(k,t_f-t)$. Therefore, at the time $t=t_f$ one has $c_a(t_f)=A^*(0)=1$ and  $\varphi(k,t_f)=\Phi^*(k,0)=0$, i.e. $\rho^{(2)}(t_f)=\rho^{(1)}(0)$. This means that at the time $t_f$ we are in the same initial state as in the previous case, and thus for $t>t_f$ one clearly has $c_a(t)=A(t-t_f)$ and $\varphi(k,t)=\Phi(k,t-t_f)$, i.e. $D(\rho_S^{(2)}(t))=D(\rho_S^{(1)}(t-t_f))$. 
Basically, starting from the initial entangled state (8) in the initial time interval $0<t<t_f$ there is a backflow from the photon field into the atom, i.e. there is absorption of radiation until at time $t=t_f$ the atom is exactly in the excited state. In the following time interval $t>t_f$, an irreversible atom decay with emission of a photon occurs as in the previous (canonical) case, but delayed by the time $t_f$. This dynamical behavior clearly leads to a pronounced ME when comparing the relaxation of states $\rho_S^{(1)}(t)$ and  $\rho_S^{(2)}(t)$. This is illustrated in Fig.2(b) for the coupled resonator optical waveguide setup, where curve  2 depicts the numerically-computed decay curve $D(\rho_S^{(2)}(t))$, which should be compared to the canonical decay curve 1.
 \begin{figure}
\includegraphics[width=8.5 cm]{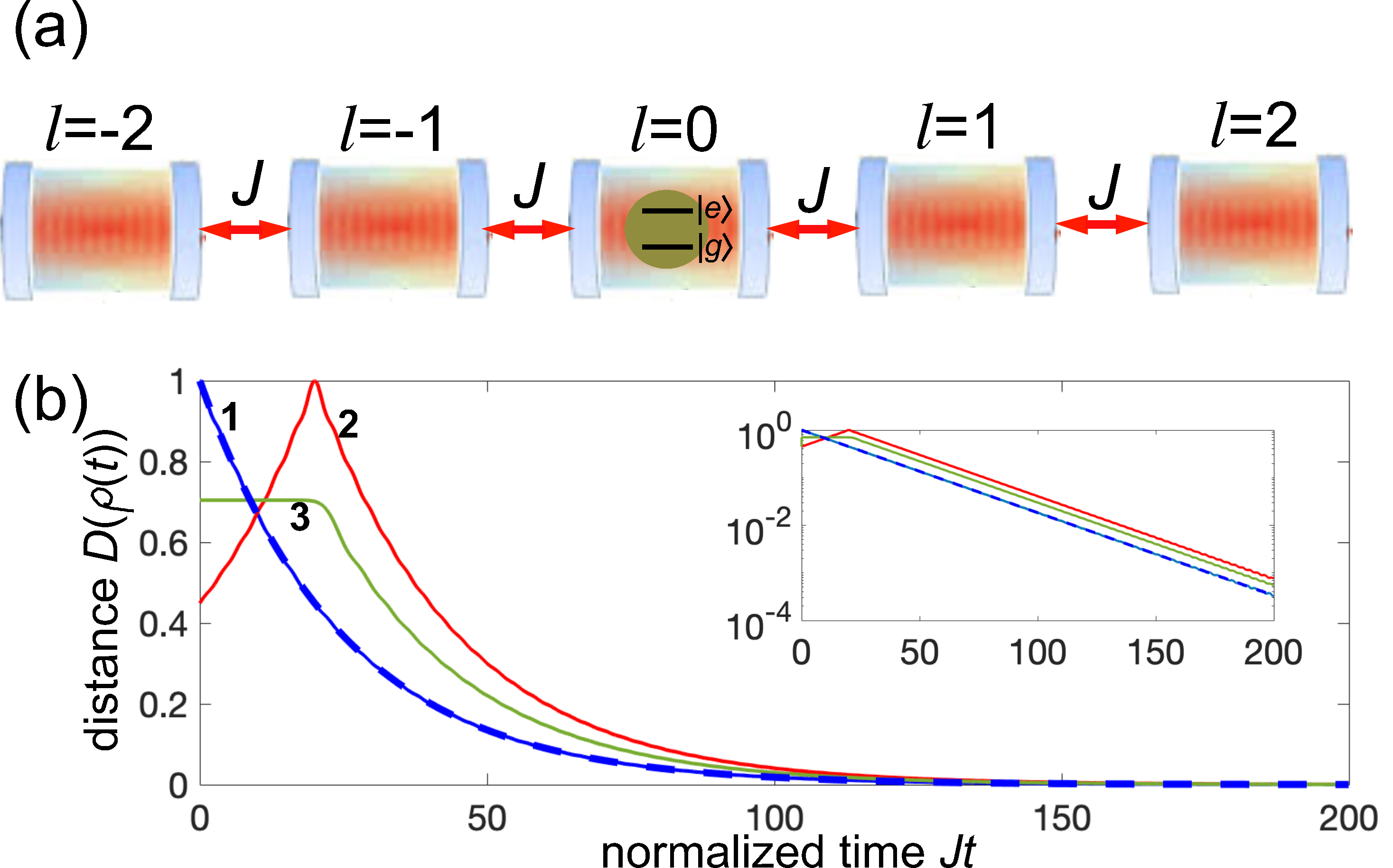}
\caption{(a) Schematic of an array of coupled optical cavities with a two level atom placed in the resonator of index $l=0$. $J$ is the hopping rate of photons between adjacent resonators of the array. (b)  Numerically-computed temporal behavior of the trace distance $D(\rho_S(t))$, which is equal to the survival probability $|c_a(t)|^2$ to find the atom  in the excited state $|e \rangle$, corresponding to three different initial conditions $\rho(0)$ of the atom-photon field density operator in the single excitation sector and for parameter values $g_0/J=0.2$ and $\omega_0=\omega_c$. Curve 1 corresponds to the canonical initial condition $\rho(0)=|e \rangle \langle e|  \otimes |0 \rangle \langle 0|$, with the atom initially in the excited state and the photon field in the vacuum state $|0 \rangle$. The decay curve $|c_a(t)|^2=|A(t)|^2$ is very well fitted by an exponential curve $\exp(- \Gamma t)$, with the spontaneous emission rate $\Gamma=g_0^2/J$ given by the Fermi golden rule. 
Curve 2 corresponds to the initial entangled atom-photon  state $\rho(0)=| \psi(0) \rangle \otimes \langle \psi(0)|$, with $| \psi(0) \rangle$ given by Eq.(8) with $t_f=20/J$. Finally, curve 3 corresponds to the the initial entangled atom-photon  state $\rho(0)=| \psi(0) \rangle \otimes \langle \psi(0)|$, with $| \psi(0) \rangle$ given by Eqs.(9) and (10) with  $L=20$ The inset in (b) displays the three decay curves on a log time scale, clearly showing that the long-time decay behavior is the same for the three initial conditions.}
\end{figure}

The observation of the ME predicted in the above analysis requires a careful preparation of the system-bath  in an initial entangled state based on time reversal dynamics, which could be in principle implemented in lattice systems using quantum Loschmidt echo and quantum mirror methods (see e.g. \cite{tilt1,tilt2,tilt3,tilt4,tilt5,tilt6,tilt7}). For example, using photonic lattice setups to emulate spontaneous emission of a quantum emitter in a waveguide of coupled resonators \cite{emu1,emu2,emu3,emu4}, the time reversal procedure required to prepare the initial system-bath entangled state can be readily implemented starting from the canonical initial condition, leaving the full system freely evolving for a time interval $t_f$ and then applying $\pi$ lumped phase slips at alternating lattice sites (an experimental demonstration of such a time-reversal procedure has been reported in \cite{emu5}).
  However, we stress that the emergence of the ME from system-bath entanglement is a general feature, not specific to the special time-reversed initial condition $\rho^{(2)}(0)$ considered in the previous analysis as an illustrative example. In fact, the ME is expected to be observable rather generally when condition (2) is not satisfied in the early times of the dynamics, with transient backflow from the photon field into the atom which can transiently suppress or even reverse the spontaneous emission process.  As shown in Appendix B, this happens for a very broad class of initial atom-photon entangled states $\rho^{(3)}(0)$: for such states, the distance $D(\rho_S^{(3)}(t))$ from the equilibrium state  can transiently increase in time, while in the long time limit, $t \rightarrow \infty$,  $D(\rho_S^{(3)}(t))$  displays the same exponential decay law as $D(\rho_S^{(1)}(t))$, i.e. of the canonical initial condition with the atom in the excited state and the photon field in vacuum state.  For example, for the coupled resonator optical waveguide of Fig.2(a) with $\omega_c=\omega_0$, let us consider the decay dynamics corresponding to the initial condition $\rho^{(3)}(0)= | \psi(0) \rangle \langle \psi(0)|$ with $|\psi(0) \rangle$ given by (in Wannier basis) 
\begin{equation}
|\psi_(0)=c_a(0) |e \rangle \otimes |0 \rangle+ \sum_l Q_l \hat{c}^{\dag}_l |0 \rangle \otimes |g \rangle
\end{equation}
with only ($2L+1)$ nonvanishing $Q_l$ amplitudes, namely $Q_0=-c_a(0) J/g_0$, $Q_2=Q_{-2}=-Q_0$, $Q_4=Q_{-4}=Q_0$, ..., $Q_{2L}=Q_{-2L}=(-1)^{L} Q_0$ and
\begin{equation}
|c_a(0)|^2= \frac{1}{1+(2L+1) J^2/g_0^2}
\end{equation}
for normalization. In the large $L$ limit, this state is a quasi-equilibrium state of the full atom-photon system since $| \psi(0) \rangle$ is an eigenstate of the full Hamiltonian $H$ for $L \rightarrow \infty$, as one can readily check: in this limit, we thus expect the dynamics to be frozen and the atom decay suppressed. For a finite $L$, the suppression of spontaneous emission is only transient and holds for the time interval $0<t<\Delta t$ with $\Delta t \sim 2L/v_g=L/J$. In other words, the initial atom-photon state given by Eqs.(9) and (10) delays the spontaneous emission process by a time interval $\Delta t$. An example of the exact numerically-computed decay behavior of $|c_a(t)|^2$ for this initial entangled state is shown by curve 3 in Fig.2(b), clearly indicating the onset of ME when compared with the decay of the canonical initial state (curve 1 in the figure).

In conclusion, we have introduced a novel type of quantum Mpemba effect that originates from initial system-bath entanglement, which plays a crucial role in shaping the early stages of thermalization. This mechanism is fundamentally distinct from previously studied Mpemba scenarios-- most notably the strong Mpemba effect-- which are typically associated with enhanced or accelerated relaxation for specific initial system states. In contrast, the effect described here arises not from modified relaxation rates intrinsic to the system, but from pre-existing quantum correlations with the environment, leading  to a characteristic  delayed thermalization dynamics. We illustrated this phenomenon through the spontaneous emission of a two-level atom coupled to a photonic waveguide at zero temperature. When the atom and photon field are initially entangled in a time-reversed decayed state, the emission process is significantly delayed, offering a clear and experimentally relevant manifestation of the entanglement-induced Mpemba effect. Our findings open the door to identifying new classes of Mpemba-like effects in quantum systems, fundamentally shaped by initial quantum correlations. Beyond deepening our understanding of quantum thermalization, these insights hold promise for applications in quantum thermodynamics, quantum electrodynamics, and emerging quantum technologies, where precise manipulation of energy flow and thermalization is essential.

\begin{acknowledgments}
The author acknowledges the Spanish State Research Agency, through
the Severo Ochoa and Maria de Maeztu Program for Centers 
and Units of Excellence in R\&D (Grant No. MDM-2017-
0711).
\end{acknowledgments}

\section*{Conflict of Interest Statement}
The author has no conflicts to disclose.

\section*{Data Availability Statement}
The data that support the findings of this study are available within the article.

\appendix

{\color{black}
\section{System dynamics under time-reversal symmetry}
Let us assume that the full system-bath Hamiltonian $H$ has time-reversal symmetry. This symmetry is expressed as $H \mathcal{K}=\mathcal{K}H$, where $\mathcal{K}$ is the anti-unitary time-reversal operator that we assume to be the complex conjugation operation \cite{revers}. In this Appendix we show that for any initial state of the composite system described by the density operator $\rho(0)=\rho_S(0) \otimes \rho_B$ satisfying the condition $\mathcal{K} \rho_S(0) \mathcal{K}= \rho_S(0)$, if at some time $t_f$ the state $\rho_S(t_f)={\rm Tr}_B (U_{t_f} \rho(0) U_{t_f}^{\dag})$ is close to the equilibrium state $\rho_E$, than the state $\rho_S^{(2)}={\rm Tr}_B(\rho(-t_f))= {\rm Tr}_B (U_{-t_f} \rho(0) U_{-t_f}^{\dag})$ is close to the equilibrium state as well.\\
In fact, taking into account that under time reversal symmetry $\mathcal{K}i \mathcal{K}=-i$, it readily follows that $U_{-t}= \mathcal{K} U_t \mathcal{K}$, where $U_t=\exp(-iHt)$ is the unitary time evolution operator of the composite system. Since $U^{\dag}_t=U_{-t}$, one then obtains
\begin{equation}
\rho(-t)=U_{-t} \rho(0) U_{t}=\mathcal{K}U_t \mathcal{K} \rho(0) \mathcal{K} U^{\dag}_t \mathcal{K}
\end{equation}
Assuming that the initial state $\rho(0)$ is the product state $\rho(0)=\rho_S(0) \otimes \rho_B$, with $\mathcal{K} \rho_B \mathcal{K}= \rho_B$ 
and $\mathcal{K} \rho_S(0) \mathcal{K}= \rho_S(0)$, one has
\begin{equation}
\rho(-t)=\mathcal{K}U_t \rho(0) U^{\dag}_t \mathcal{K}
\end{equation}
and thus
\begin{equation}
\rho_S^{(2)}={\rm Tr}_B \left\{ \mathcal{K}U_{t_f} \rho(0) U^{\dag}_{t_f} \mathcal{K} \right\}= \mathcal{K} {\rm Tr}_B \left\{U_{t_f} \rho(0) U^{\dag}_{t_f}  \right\} \mathcal{K}.
\end{equation}
From Eq.(A.3) one then obtains
\begin{equation}
\rho_S^{(2)}= \mathcal{K} \rho_S(t_f) \mathcal{K}
\end{equation}
i.e.
\begin{equation}
\langle n | \rho_S^{(2)} | m \rangle= \left( \langle n | \rho_S(t_f) | m \rangle \right)^*=\langle m | \rho_S(t_f) | n \rangle 
\end{equation}
where $\{ |n \rangle \}$ is the energy eigenbasis of $H_S$. From Eq.(A.5) it then follows that, if $\rho_S(t_f)$ is close to the equilibrium state $\rho_E$, which is diagonal in the $\{ | n \rangle \}$ basis, then also $\rho_S^{(2)}$ is close to the equilibrium state. Finally, we note that, since $\mathcal{K}$ is the complex conjugation operator, the condition $\mathcal{K} \rho_S(0) \mathcal{K}= \rho_S(0)$ on the initial state $\rho_S(0)$ of the system is equivalent to state that the matrix $\langle n | \rho_S(0) | m \rangle$ is symmetric, besides to be Hermitian.

\section{Atom-photon field dynamics: exact analysis}
In the single excitation sector $N=1$, the  exact atom-photon dynamics is described by the coupled equations (6) and (7) given in the main text, which should be solved with assigned initial atom-photon state, i.e. $c_a(0)$ and $\varphi(k,0) \equiv \varphi_0(k)$. After letting  $\Omega(k)=\omega(k)-\omega_0$, the formal solution of Eq.(7) reads
\begin{equation}
\varphi(k,t) =\left\{ \varphi_0(k)-ig^*(k) \int_0^t dt' c_a(t') \exp[i \Omega(k) t'] \right\} \exp[-i\Omega(k)t].
\end{equation}
Substitution of Eq.(B1) into Eq.(6) yields the following integro-differential equation for $c_a(t)$ 
\begin{equation}
\frac{dc_a}{dt}=F(t)- \int_0^t dt' \mathcal{G}(t-t') c_a(t')
\end{equation}
where the memory function $\mathcal{G}(\tau)$ and forcing term $F(t)$ are given by
\begin{eqnarray}
\mathcal{G}(\tau) & = & \int dk |g(k)|^2 \exp[-i \Omega(k) \tau] \\
F(t) & = & -i \int dk g(k) \varphi_0(k) \exp[-i \Omega(k) t]. 
\end{eqnarray}
We note that, from the stationary phase method it follows that  both $\mathcal{G}(\tau)$ and $F(t)$ decay to zero as $\tau, t \rightarrow \infty$. In particular, the characteristic decay time $\tau_m$ of $\mathcal{G}(\tau)$ defines the memory time of the non-Markovian dynamics.\\  
Let us first consider the canonical initial condition, with the atom initially in the excited state and the photon field in the vacuum state, corresponding to $c_a(0)=1$ and $\varphi_0(k)=0$. The solution $c_a(t)$ for such a canonical initial condition is indicated by $A(t)$. Since the forcing term $F(t)$ in Eq.(B2) vanishes and clearly $|A(t)|^2 \leq |c_a(0)|^2=1$, the condition (2) is not violated. For a featureless continuum and in the absence of atom-photon bound states, $A(t) \rightarrow 0$ as $t \rightarrow \infty$. As is well known, in the Markovian limit $\tau_m \rightarrow 0$ the decay law $|A(t)|^2$ becomes exponential with a decay rate $\Gamma$ given by the Fermi golden rule. In fact, for $\tau_m \rightarrow 0$ the intergo-differential equation (B2) with $F(t)=0$ can be replaced by the differential equation
\begin{equation}
\frac{d A}{dt} \simeq - \left( \frac{\Gamma}{2}+i \Delta \right) A(t)
\end{equation}
where the decay rate $\Gamma$ and frequency Lamb shift $\Delta$ are obtained from the relation
\begin{equation}
\frac{\Gamma}{2}+ i \Delta = \int_{0}^{\infty} d \tau \mathcal{G}( \tau)=  \int dk |g(k)|^2 \int_0^{\infty} d \tau \exp[-i \Omega(k) \tau] \
\end{equation}
Taking into account that
\begin{equation}
\int_0^{\infty} d \tau \exp[-i\Omega(k) \tau]= \pi \delta(\Omega(k)) - i \mathcal{P} \left( \frac{1}{\Omega(k)} \right),
\end{equation}
where $\mathcal{P}$ denotes the principal value,
one obtains
\begin{eqnarray}
\Gamma & = & 2 \pi \int dk |g(k) |^2 \delta(\omega(k)-\omega_0) \\
\Delta & = & \mathcal{P} \int dk \frac{|g(k)|^2}{\omega_0-\omega(k)}.
\end{eqnarray}
Indicating by $\pm k_0$ the two opposite values of the wave number of the electromagnetic modes such that $\omega( \pm k_0)=\omega_0$ and letting $v_g=|(d \omega/ dk)_{\pm k_0}|$, in the integral on the right hand side of Eq.(B8) we have two contributions, coming from $k= \pm k_0$. Assuming $g(-k_0)=g(k_0)$, which holds for a non-chiral waveguide, the two contributions are the same and one finally obtains
 \begin{equation}
\Gamma=\frac{4 \pi |g(k_0)|^2}{v_g}. 
\end{equation}
Finally, let us consider the more general case of an initial atom-photon entangled state, i.e. let us assume $c_a(0), \varphi_0(k) \neq 0$, corresponding to a non-vanishing forcing term $F(t)$ in Eq.(B2).
Since $(d D(\rho_S(t)) /dt)=(d|c_a(t)|^2 /dt)=2 {\rm Re} \{ c^*_a(t) (dc_a/dt)\}$, from Eq.(B2) it readily follows that at initial time $t=0$ the slope of the distance $D(\rho_S(t))$ is given by
\begin{equation}
\left( \frac{d D(\rho_S(t))}{dt} \right)_{t=0}= 2 {\rm Re} \left\{ c_a^*(0) F(0) \right\}
\end{equation}
i.e. 
\begin{equation}
\left( \frac{d D(\rho_S(t))}{dt} \right)_{t=0}=2 {\rm Im} \left\{ c_a^*(0) \int dk g(k) \varphi_0(k) \right\}
\end{equation}
where we used Eq.(B4). From Eq.(B12) one sees that there is a broad class of initial entangled states for which the slope of the distance $D(\rho_S(t))$ at initial time can be positive, resulting in the violation of Eq.(2). On the other hand, provided that driving term $F(t)$ vanishes fast enough as $t \rightarrow \infty$, the long time dynamics of $c_a(t)$ is not much influenced by the initial atom-photon entanglement in the weak coupling regime, i.e. the final decay rate of $c_a(t)$ is the one given by the Fermi golden rule.


\end{document}